\documentclass[aps,prl,reprint,superscriptaddress]{revtex4-1}
\usepackage{graphicx}
\usepackage{dcolumn}
\usepackage{bm}
\usepackage{graphicx}
\usepackage{dcolumn}
\usepackage{bm}
\usepackage{subfigure}
\usepackage{color}
\usepackage{amsmath}
\usepackage{epstopdf}

\begin{document}


\renewcommand{\thetable}{\arabic{table}}

\newcommand{\change}[1]{\textcolor{red}{#1}} 
\newcommand{\BU}{Department of Mechanical Engineering, Boston University, Boston, Massachusetts 02215, USA}

\newcommand{\VA}{Department of Medicine, Boston University School of Medicine, Boston, Massachusetts 02118, USA}


\title{All-electrical monitoring of bacterial antibiotic susceptibility in a microfluidic device}

\author{Yichao Yang}
\affiliation{\BU}

\author{Kalpana Gupta}
\affiliation{\VA}

\author{Kamil L. Ekinci}
\email[Electronic mail: ]{ekinci@bu.edu}
\affiliation{\BU}

\date{\today}

\begin{abstract}

The lack of rapid antibiotic susceptibility tests adversely affects the treatment of bacterial infections and contributes to  increased prevalence of multidrug resistant bacteria. Here, we describe an all-electrical approach that allows for ultra-sensitive  measurement  of growth signals from only tens of bacteria in a microfluidic device. Our device is essentially a set of microfluidic channels, each with a nano-constriction at one end and cross-sectional dimensions close to that of a single bacterium.  Flowing a liquid bacteria sample (e.g., urine) through the microchannels rapidly traps the bacteria in the device, allowing for subsequent incubation in drugs. We measure the electrical resistance of the microchannels, which increases (or decreases) in proportion to the number of bacteria  in the microchannels. The method and device allow for rapid antibiotic susceptibility tests in about two hours. Further, the short-time  fluctuations in the electrical resistance during an antibiotic susceptibility test are correlated with the morphological changes of bacteria caused by the antibiotic. In contrast to other electrical approaches, the underlying geometric blockage effect provides a robust and sensitive signal, which is straightforward to interpret without electrical models. The approach also obviates the need for a high-resolution microscope and other complex equipment, making it potentially usable in resource-limited settings.


\end{abstract}

\maketitle

\section{Introduction}

Multidrug resistant bacteria pose an increasingly serious threat to global public health \cite{brown2016antibacterial}. While drug resistance in bacteria occurs naturally due to random genetic mutations and genetic exchanges between strains and species, it is accelerated partly because of inappropriate antibiotic use \cite{blair2015molecular}. Strategies, such as rapid point-of-care antibiotic susceptibility testing, can facilitate targeted antibiotic treatments and impede the spread of antibiotic resistance \cite{ataee2012method, jenkins2012current, van2018developmental}.  However, standard antibiotic susceptibility tests (ASTs) suffer from a lengthy cell culture step and take 24-48 hours to complete \cite{balouiri2016methods, khan2019current}. Given the risks associated with delayed therapy, physicians typically have little choice but to empirically prescribe broad-spectrum antibiotics while waiting for the microbiological analysis \cite{gonzalez2015effect, copp2011national}. The development  of rapid ASTs would improve morbidity and mortality and could help reduce the prevalence of multidrug resistant bacteria \cite{li2018bacteria}. 

The ``gold standard” ASTs are phenotypic and measure the growth of bacteria in the presence of antibiotics on solid agar plates or in liquid solutions. After incubation for 24-48 hours, the susceptibility of the bacterial strain can be determined from the growth size and patterns on the plate or the optical density (OD) of the liquid solution  \cite{jenkins2012current}. Polymerase chain reaction (PCR) provides the quintessential genotypic AST \cite{rajivgandhi2018detection}. PCR directly detects the resistance gene(s) from a very small bacteria sample and hence is quite rapid. However, it still has limited utility, because only a few resistance genes are firmly associated with phenotypic antibiotic resistance and  newly-acquired resistance mechanisms may not be detectable \cite{hughes2017environmental}.  

Given the limitations of mainstay ASTs, there is  a significant push for developing novel methods that can inform on bacterial resistance at early stages of cell growth. These novel and emerging ASTs typically employ microfluidics and microdevices because these devices allow for effective sample use and are sensitive to small signals \cite{dai2016microfluidics}.  State-of-the-art approaches isolating bacteria in nanodroplets \cite{kang2019ultrafast,boedicker2008detecting}, on microbeads \cite{wang2018label}, inside microfluidic channels \cite{baltekin2017antibiotic,li2019adaptable,mannoor2010electrical,chen2016fast,besant2015rapid}, and on and inside micromechanical resonators \cite{stupar2017nanomechanical} have all allowed testing on a few cells and even single cells. These approaches  involve a variety of transduction mechanisms to access the response of bacteria to antibiotics, including high-resolution imaging \cite{kang2019ultrafast,boedicker2008detecting,baltekin2017antibiotic,li2019adaptable}, mechanical  \cite{wang2018label,stupar2017nanomechanical}, impedance \cite{mannoor2010electrical}, and  electrochemical sensing \cite{chen2016fast,besant2015rapid}.  More recently, high-resolution imaging of growth of bacteria trapped in microchannels \cite{baltekin2017antibiotic,li2019adaptable} have allowed for ASTs in under an hour \cite{baltekin2017antibiotic}. While ingenious, each method comes with some drawbacks \cite{behera2019emerging}, and it remains to be seen whether or not any  will achieve sufficient robustness needed for routine clinical practice.


\begin{figure*}[t]
\centering
\includegraphics[width=6 in]{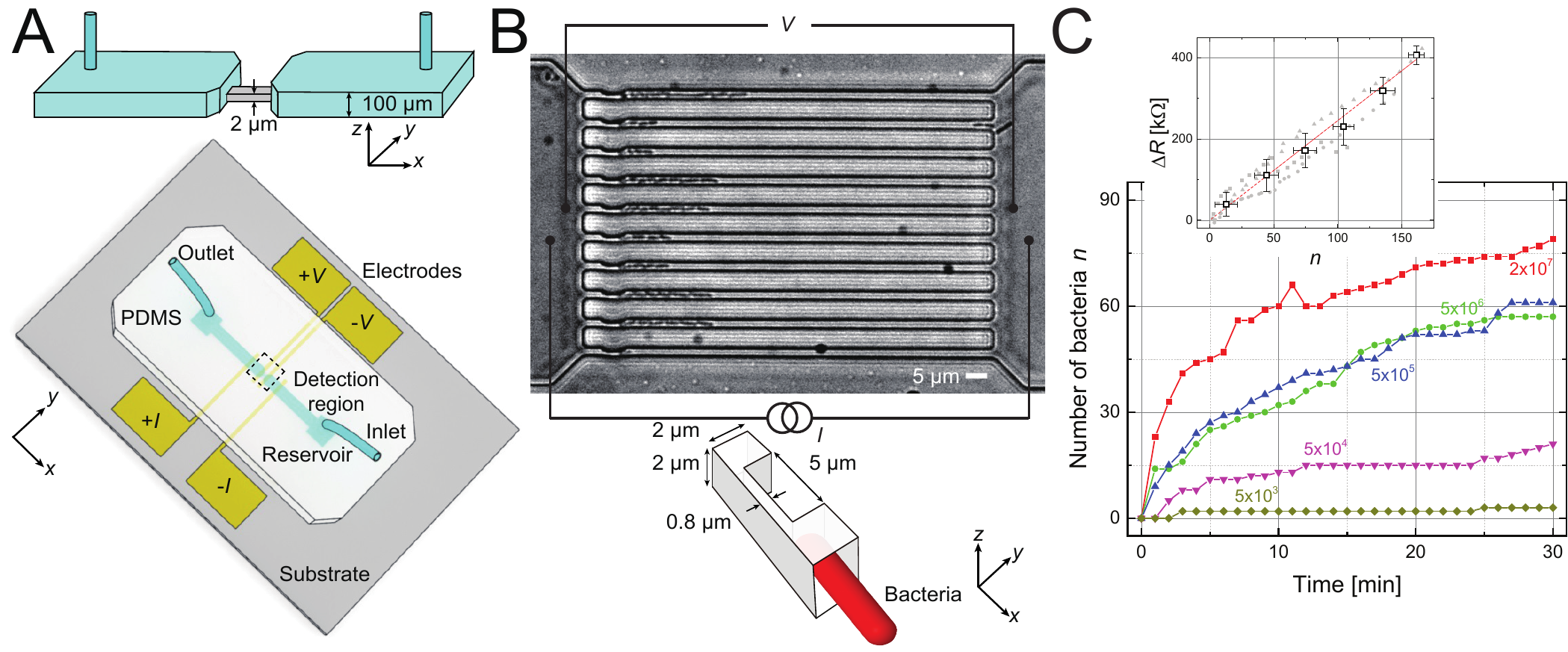}
\caption{Microfluidic device and its principle of operation. (\textit{A}) Schematic of the device. The PDMS slab embedded with a two-layer microfluidic channel (inset) is bonded onto a glass substrate with deposited thin film electrodes. At the center is the detection region, which features an array of ten microchannels (2-$\mu$m height and 2-$\mu$m width) in parallel. These central microchannels are connected to two reservoirs via 100-$\mu$m-height macrochannels; the two reservoirs are connected to  sample lines. (\textit{B}) Microscope image ($63\times$) of trapped bacteria (\textit{K. pneumoniae}) in the microchannels. Scale bar is 5 $\mu$m. Bottom illustration shows the constriction for capturing bacteria.  A four-wire electrical resistance measurement is used. Growth and morphological changes of  bacteria in the microchannel alter the effective electrical resistance of the microchannel.  (\textit{C}) Number of trapped bacteria (\textit{K. pneumoniae}) in the microchannels as a function of sample loading time for cultures with different cell concentrations. Inset shows the electrical resistance change as a function of the number of bacteria in the microchannel from three nominally identical devices. The linear fit gives the resistance change per added bacterium of $\sim 2.5~\rm k \Omega$; the large data points correspond to binned average values.}
\label{fig:figure1}
\end{figure*}

Our method and device build on the positive attributes of recent approaches and address some of their shortcomings. As in earlier work \cite{baltekin2017antibiotic,li2019adaptable}, we trap and incubate cells in a microfluidic channel; our measurement, however, is entirely electrical. The  effect underlying the bacterial growth signal in our device is simple geometric blockage:  as bacteria grow (or die) in the microchannel, the channel resistance to electrical current increases (decreases). The  change in the number of bacteria in the device, therefore, is directly proportional to the measured  resistance change, and  is available without fits to multi-parameter circuit models  \cite{yang2008electrical}. The device can directly be used with urine and probably other bodily fluids, provided that the fluids contain ions. Another interesting and useful attribute of the  approach is that it provides electrical clues on how bacteria respond to antibiotics.  We observe  different short-time fluctuation patterns in  electrical signals coming from bacteria incubated in  bacteriostatic and bactericidal antibiotics, suggesting that morphological changes are also encoded into the electrical signals. In cases where both growth and morphological analysis are required, this unique feature may be useful \cite{choi2014rapid, pitruzzello2019multiparameter, khan2019progress}. We re-emphasize that microscopy is not required in our approach; at the current stage of development, it is used only as a validation tool.

\section{Results}

\subsection*{Device Design and Loading}

 The design and basic principle of operation  of the microfluidic device is shown in Fig. \ref{fig:figure1}. The device is essentially a continuous polydimethylsiloxane (PDMS)  channel on a glass substrate with thin metal film electrodes. At the center, the channel tapers down into ten smaller microchannels each with linear dimensions $l\times w\times h \approx 100\times 2 \times 2 ~\mu \rm m^3$. On one end of each of these parallel microchannels, a physical constriction ($l\times w\times h \approx 5\times 0.8 \times 2 ~\mu \rm m^3$) is fabricated for trapping bacteria from a flowing sample (Fig. \ref{fig:figure1}\textit{B}). During operation, a pressure-driven flow of a bacteria solution is established through the microchannels from  the inlet to the outlet (Fig. \ref{fig:figure1}\textit{A} and \textit{B}). The bacteria in the solution cannot pass through the constriction and are trapped as shown in  the optical microscope image in Fig. \ref{fig:figure1}\textit{B}.  In the experiments, the  electrical resistance of the microchannel region is monitored using a  4-wire measurement (Fig.~\ref{fig:figure1}\textit{B}). When the microchannels are filled with just media with no bacteria, i.e., empty, their typical resistances are $R_{em} \approx 3 ~\rm M \Omega \pm 30 ~\rm k\Omega$.
 
At the start of each experiment, the bacteria sample is loaded into the microfluidic device from the inlet by keeping the inlet at a pressure  $\Delta p \sim10$ kPa above the outlet. Fig. \ref{fig:figure1}\textit{C} shows the  number of trapped cells as a function of time for \textit{Klebsiella pneumoniae} suspensions at different bacteria concentrations in the range from $5\times 10^{3}$ to $2\times 10^{7}$ CFU/mL. Approximately 60 bacteria are captured in $\sim 30$ min at 5$\times 10^{5}$ CFU/mL, which is close to the cell density in the urine of a UTI patient \cite{duell2012innate}. At the most dilute bacteria  concentration (5$\times 10^{3}$), we only trapped a few bacteria over the 30 min period. The trapping efficiency at low concentration could be improved  by increasing $\Delta p$ or the number of microchannels. 

The inset of Fig. \ref{fig:figure1}\textit{C} shows the electrical signal, as bacteria number increases in the microchannels. Here, we show the resistance increase $\Delta R$ from the empty state,  as a function of the number $n$ of bacteria  in the microchannels from three separate experiments using different microchannels with identical nominal linear dimensions. The increase in $\Delta R$ with $n$ can be understood in simple terms. The microchannel filled with the media is essentially an electrical conductor due to the ions in the buffer. The bacteria in the microchannel ``clog'' the microchannel and reduce the effective cross-section, thereby increasing the resistance. The data follow a linear trend, with a resistance change of $\Delta R^{(KP)}_1 \approx 2.5 \pm 0.3 ~\rm k\Omega$ per bacterium (\textit{K. pneumoniae}) added.  Similar experiments (SI Appendix, Fig. S3) give  $\Delta R^{(EC)}_1 \approx 3.7 \pm 0.3 ~\rm k\Omega$ and $\Delta R^{(SS)}_1 \approx 3.5 \pm 1.1 ~\rm k\Omega$ for \textit{Escherichia coli} and  \textit{Staphylococcus saprophyticus}, respectively.  An estimate using simple geometric arguments for \textit{K. pneumoniae} provides $\Delta R_1 \sim 1.5~\rm k\Omega$, not far from the measured value (SI Appendix, Supplemental Materials and Methods). These $\Delta R_1$ values provide  a good calibration for the experiments and allow us to the estimate that $\sim 20$ cells are needed to to perform a conclusive antibiotic susceptibility test due to the long-term electrical drifts. 

Since we do not use very high pressures during loading, bacteria occasionally accumulate at other locations in the device, particularly at lithographical edges and at the entrances of microchannels. When this happens, the measured resistances to correspond to larger bacteria numbers than  counted from microscope image. The electrical signal is  quite robust against such non-ideal occurrences. First, the geometry ensures that the largest resistance signals come from the central detection region (SI Appendix, Supplemental Materials and Methods). Second, all bacteria inside the device, regardless of where they are, generate coherent electrical signals of cell growth or cell death. Third, any contaminants that partially block the device and do not change over time just result in time-independent background signals.

\subsection*{Electrical Monitoring of Bacterial Growth}

\begin{figure}[t]
\centering
\includegraphics[width=3 in]{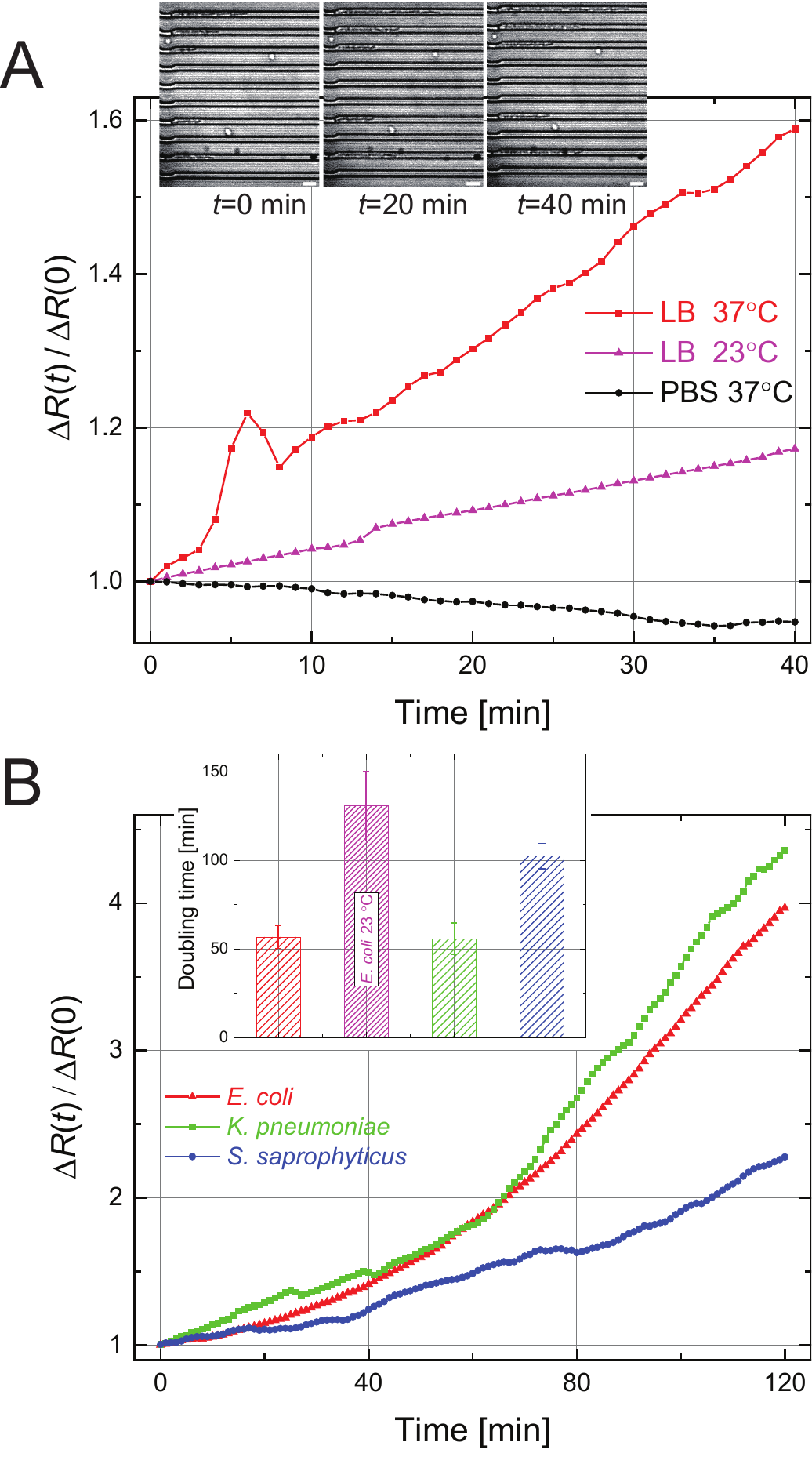}
\caption{ Electrical detection of bacteria growth. (\textit{A}) Growth curves  for \textit{E. coli} in PBS at $37^{\circ}$C (black) and LB broth at $23^{\circ}$C (magenta) and $37^{\circ}$C (red). Inset shows optical images of \textit{E. coli}  in the microchannels in LB broth at $37^{\circ}$C at different points in time. The  scale bars are 5 $\mu$m. (\textit{B}) Growth curves for  \textit{E. coli} (red), \textit{K. pneumoniae} (green), and \textit{S. saprophyticus} (blue) at $37^{\circ}$C in LB broth. Each data trace is the average of three independent experiments. (Error bars for the growth data are shown in Fig. \ref{fig:figure3}.) Inset shows the doubling time. Error bars represent standard deviations.}
\label{fig:figure2}
\end{figure}

We first perform an electrical measurement of bacteria growth. We record the device resistance $R(t)$ as a function of time. We then determine the normalized time-dependent resistance change defined as ${{\Delta R(t)} \over {\Delta R(0)}} = {{R(t) - {R_{em}}} \over {R(0) - {R_{em}}}}$, where $R(0)$ and $R_{em}$ respectively are the device resistance right after loading ($t=0$) and without bacteria (empty). From Fig. \ref{fig:figure1}\textit{C} inset above,  we expect that ${{\Delta R (t)}\over{\Delta R (0)}} \approx {{n (t)}\over{n(0)}}$, where $n(t)$ is the number of bacteria in the microchannels. Fig. \ref{fig:figure2}\textit{A} shows the normalized resistance  for motile \textit{E. coli} as a function of time obtained in Phosphate-buffered saline (PBS) at $37^{\circ}$C and Luria-Bertani (LB) broth at $23^{\circ}$C and  $37^{\circ}$C.  After bacteria are loaded and during the measurement, $\Delta p \approx 0.5$ kPa is applied to maintain a constant flow of nutrients. Shown in Fig. \ref{fig:figure2}\textit{A} inset are optical microscope images of the trapped \textit{E. coli} taken at $t = 0, ~20, ~40$ mins during the electrical measurement in LB broth at $37^{\circ}$C (also see SI Appendix, Movie S1).


Fig. \ref{fig:figure2}\textit{B} shows similar growth curves for  gram-positive and gram-negative bacteria. \textit{K. pneumoniae} and \textit{S. saprophyticus} are non-motile and are easily trapped in the microchannels by a pressure-driven flow; \textit{E. coli} is motile but unlikely to reverse its direction and exit the tight microchannel once it enters.  The electrical resistance changes are all close to exponentials: $\frac{{\Delta R(t)}}{{\Delta R(0)}} \approx \frac{{n(t)}}{{n(0)}} \approx {e^{rt}}$,  with the growth rate $r$ providing the doubling time ${t_d} = \frac{{\ln 2}}{r}$ for each strain. The inset shows $t_{d}$ values obtained from  linear fits to the natural logarithms of the growth curves. The $t_d$ we measure are longer than those reported in the literature \cite{galli2017fast,regue2004gene,almeida1984comparison}, possibly due to the limited availability of nutrients in the microchannels \cite{yang2018analysis}.

\subsection*{Antibiotic Susceptibility Testing}

\begin{figure*}[t]
\centering
\includegraphics[width=6.75 in]{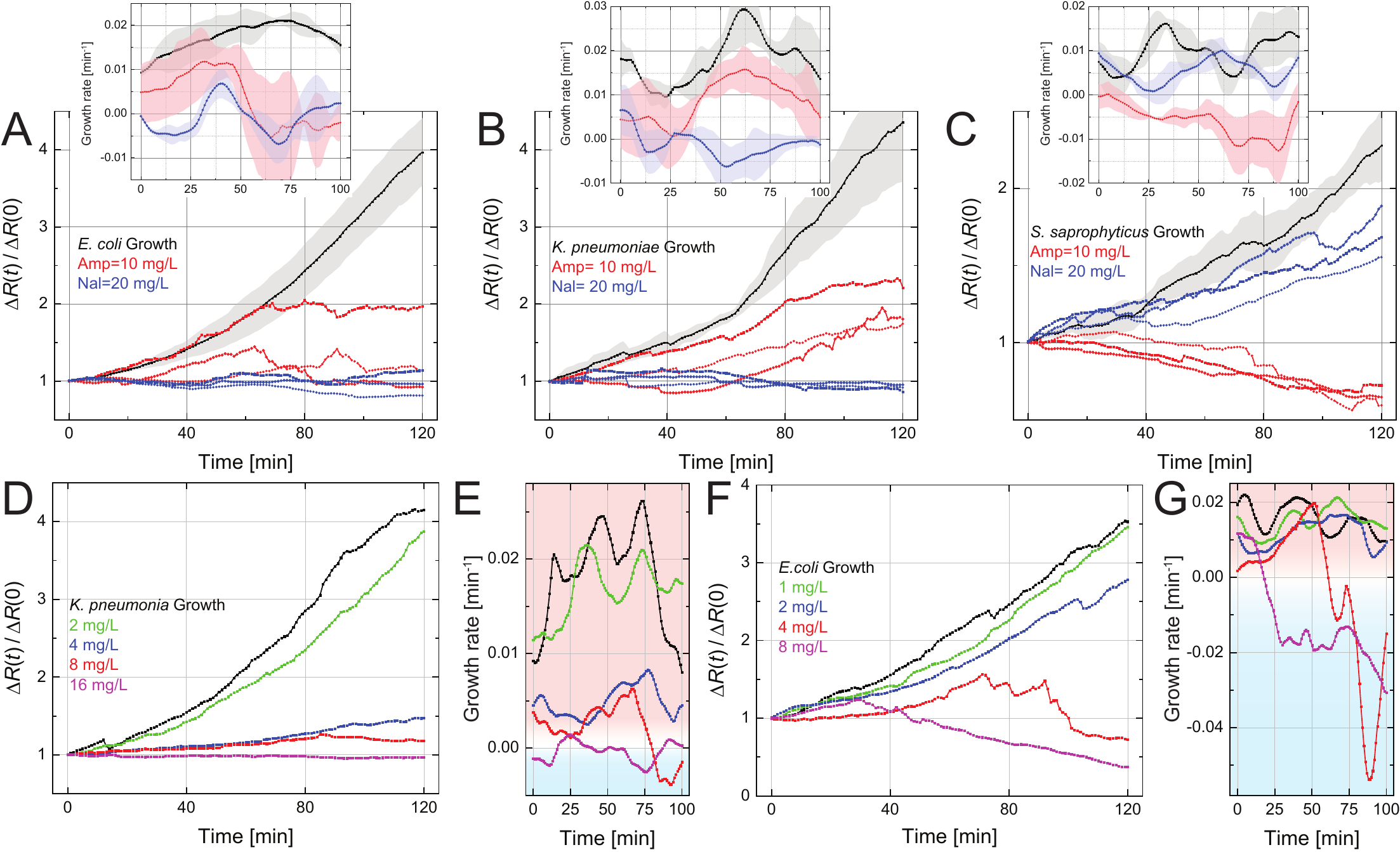}
\caption{ \textit{(A-C)} Electrical determination of the susceptibility of  \textit{E. coli} (\textit{A}),  \textit{K. pneumoniae} (\textit{B}), and   \textit{S. saprophyticus} (\textit{C}) to  ampicillin and nalidixic acid. Normalized resistance changes as a function of incubation time under different conditions are plotted for each strain. The black curve in each plot is the average  growth curve in LB broth without antibiotics at $37^{\circ}$C from Fig. \ref{fig:figure2}\textit{B}, with the shaded region showing the standard deviation; the red and blue curves show the electrical signal in LB broth with added ampicillin (10 mg/L) and nalidixic acid (20 mg/L), respectively, at $37^{\circ}$C. Each colored curve represents one independent experiment. Insets show the growth rate, $r \approx \frac{d}{{dt}}\ln \left[ {\frac{{\Delta R(t)}}{{\Delta R(0)}}} \right]$, calculated from the normalized resistance changes in the main figure; each solid line and shaded region  respectively show the average value and the standard deviation from three experiments. (\textit{D-G}) Determination of MIC.  (\textit{D}) Normalized resistance change for \textit{K. pneumoniae} as a function of incubation time in human urine with different concentrations of  nalidixic acid. (\textit{E}) Growth rates for each curve in (\textit{D}). (\textit{F}) Determination of ampicillin MIC for \textit{E. coli} in human urine.  (\textit{G}) Growth rates for each curve in (\textit{F}).}
\label{fig:figure3}
\end{figure*}

We show in Fig. \ref{fig:figure3} how our method and device can be used to determine the antibiotic susceptibility of bacteria rapidly and efficiently. We have tested bacterial response to two antibiotics with different action  mechanisms: ampicillin, a $\beta$-lactam bactericidal antibiotic, and nalidixic acid, a bacteriostatic antibiotic at low concentration. Prior to the microfluidic experiments, the susceptibility of the bacteria and the minimum inhibitory concentration (MIC) values were determined from resazurin-based microdilution ASTs (SI Appendix, Supplemental Materials and Methods and Table S1).  Each data trace in Fig. \ref{fig:figure3} was collected on a separate device. All the results are presented in terms of the normalized resistance change, ${{\Delta R (t)}\over{\Delta R (0)}}$, i.e., the approximate number of bacteria in the microchannel as a function of time normalized by the initial number of bacteria.  In each plot, the black curve shows the bacteria growth curve in LB broth with no antibiotics. The different curves show the results when bacteria are incubated in the presence of different concentrations of different antibiotics.

\begin{figure}[t]
\centering
\includegraphics[width=3 in]{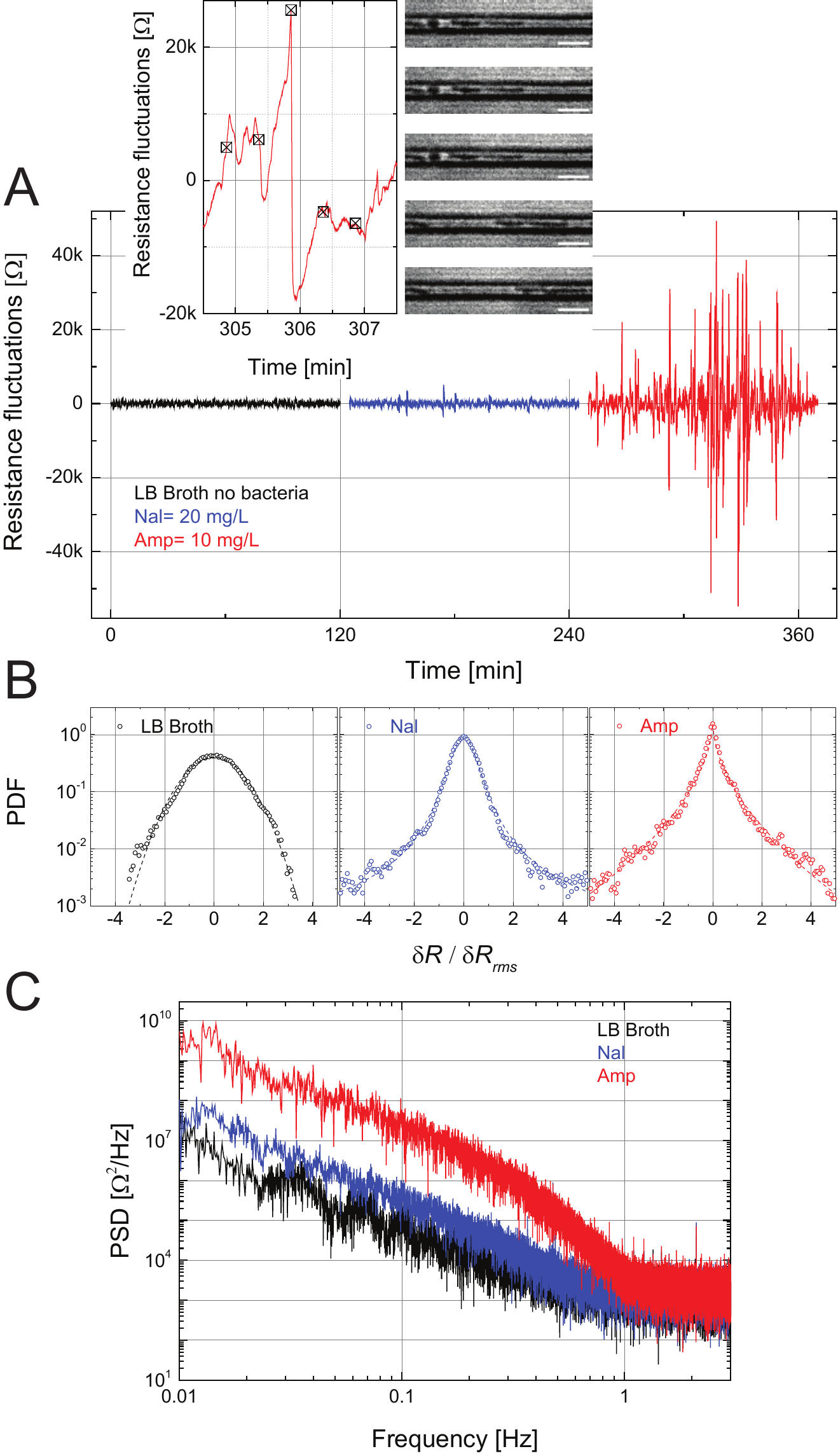}
\caption{Electrical signatures of action mechanisms of different antibiotics. (\textit{A}) Time-dependent electrical resistance fluctuations of  \textit{E. coli} in ampicillin (red trace) and nalidixic acid (blue trace). The black curve shows the baseline fluctuations in  pure LB broth with no trapped bacteria. Insets focus on a single peak along with microscope images recorded at the indicated instants, with the scale bars being 5 $\mu$m. (\textit{B})  Normalized probability density functions (PDFs) plotted in units of the rms values of the fluctuations. The black dashed line is a Gaussian. The blue dashed line is a  fit to $f(x) = A\exp ( - {{\beta {x^2}} \over {1 + \mid x{\mid ^\nu }}})$ with $A = 0.93$, $\beta = 4.65$, and  $\nu = 1.78$. The red dashed line is a fit to $f(x) = A\exp (- \beta\mid x{\mid ^\nu})$, with $A = 1.62$, $\beta = 2.85$, and  $\nu = 0.55$. (\textit{C}) The average power spectral densities (PSD) of the signals.}
\label{fig:figure4}
\end{figure}

Fig. \ref{fig:figure3}\textit{A} shows the effect of two different antibiotics, ampicillin (red curves) and nalidixic acid (blue curves), on motile \textit{E. coli}. (The black curve is the growth curve from Fig. \ref{fig:figure2}\textit{B}.) Our intial standardized ASTs  confirm that \textit{E. coli} is susceptible to both antibiotics at the indicated concentrations.  In nalidixic acid, the measured resistance is approximately constant over time (Fig. \ref{fig:figure3}\textit{A} blue curves), suggesting that the  bacteria does not grow or change in any other way. In contrast, the electrical resistance in ampicillin (Fig. \ref{fig:figure3}\textit{A} red curves) first increases but then takes a turn, staying constant or decreasing below the initial value. The behavior of the electrical resistance, without resorting to microscopy, is consistent with the fact that the cells elongate initially but cannot complete their division and eventually die. The inset of Fig. \ref{fig:figure3}\textit{A} shows the  growth rates $r$ obtained from the normalized resistance curves. Here, we compute  $\frac{d}{{dt}}\ln \left[ {\frac{{\Delta R(t)}}{{\Delta R(0)}}} \right]$ within a sliding window of 20 mins in order to reduce the numerical noise. The growth rates show that  \textit{E. coli} does not grow appreciably in either antibiotic  at the noted concentrations, suggesting the strain is susceptible to both antibiotics. A  straightforward  metric for susceptibility can be obtained by averaging the  growth rate in the second half of the test (i.e, last $\sim1$ hr). For this data set, we obtain $\bar r_G \approx 0.019~\rm min^{-1}$, $\bar r_{Amp} \approx -0.0037~\rm min^{-1}$ and $\bar r_{Nal} \approx -0.0017~\rm min^{-1}$, all averaged over three measurements. Thus,  $\bar r \le 0$ can be taken as an objective  --- albeit somewhat restrictive --- condition for susceptibility. (In Fig. ~\ref{fig:figure4} below, we  look at different aspects of the  same data for  differentiating between the action mechanisms of these antibiotics.) 

For \textit{K. pneumoniae} (Fig. \ref{fig:figure3}\textit{B}), the normalized resistance change in ampicillin (red curves) keeps increasing with incubation time, while that in nalidixic acid (blue curves) does not change at all. The inset shows the growth rates as above. The data indicate that \textit{K. pneumoniae} is resistant to ampicillin but susceptible to nalidixic acid at the indicated concentrations. We also observe that the growth rate of \textit{K. pneumoniae}  in ampicillin is lower than that with no drug.  In the case of \textit{S. saprophyticus} (Fig. \ref{fig:figure3}\textit{C}), the antibiotics cause different outcomes. 

We next determine the MICs for nalidixic acid and ampicillin using our device and method. In an effort to show the clinical relevance, we perform the MIC experiments directly in bacteria-spiked human urine mixed with LB broth.  MICs are determined within a 2-hour time window. Fig. \ref{fig:figure3}\textit{D} shows the normalized resistance change as a function of time for \textit{K. pneumoniae} in nalidixic acid at concentrations of 0, 2, 4, 8, and 16 mg/L. Increasing the concentration of nalidixic acid slows the growth down, eventually making the time derivative negative at a concentration $\lesssim16$ mg/L  (Fig. \ref{fig:figure3}\textit{D}), suggesting that 16 mg/L can safely be taken as the MIC. The corresponding growth rates in Fig. \ref{fig:figure3}\textit{E} are  negative at later times for the two highest  antibiotic concentrations.  From Fig. \ref{fig:figure3}\textit{F} and \textit{G}, we determine the MIC of ampicillin for a non-motile strain of \textit{E. coli}. The antibiotic becomes effective at a concentration  $\gtrsim 4$ mg/L but after $\sim80$ minutes of exposure.  These MIC values and our metric, $\bar r$, remain consistent  with results obtained  from standardized ASTs (SI Appendix, Table S1 and Table S2). 

\subsection*{Electrical Signatures of Antibiotic Mechanisms}

We now take a more detailed look at the data presented in Fig. \ref{fig:figure3}\textit{A} for  motile \textit{E. coli} in two antibiotics, focusing on the short-time fluctuations of the resistance. We high-pass filter the time-dependent resistance data, rejecting drifts on time scales $\gtrsim 100$ seconds. Fig. \ref{fig:figure4}\textit{A} shows samples of these resistance fluctuations as a function of time for ampicillin (red trace) and nalidixic acid (blue trace) from single two-hour measurements; the black data trace is collected in LB broth without bacteria.  There appear to be more frequent and higher-amplitude resistance fluctuations in ampicillin than nalidixic acid, with the root-mean square (rms) values being $\delta R_{amp} \approx 7.35~\rm k\Omega$ and $\delta R_{nal} \approx 0.93 ~\rm k\Omega$  during the  two-hour measurement ($\delta R_{LB} \approx 0.52 ~\rm k\Omega$). The inset on the left shows the fluctuating signal in ampicillin on the timescale of a single fluctuation. Simultaneous time-lapse microscope images (right inset) have allowed us to speculate about the source of this particular fluctuation in ampicillin. The images show that a bacterium in one of the 10 microchannels undergo a rapid burst at roughly the same time as the disappearance of the sharp electrical peak. We speculate that the swelling of the bacteria increases the resistance before the burst, and the rapid burst gives rise to the sudden resistance drop (SI Appendix, Movie S2). The microscope images show that bacteria to the right of the bursting bacterium are displaced even further and some residue remains in the microchannel after the burst. Bacteria in the other microchannels stay unchanged during this time interval.  We have not noticed many similar cell bursting events in the 2-hour time-lapse images of bacteria in nalidixic acid, which only inhibits cell division.

To provide more quantitative insight into antibiotic mechanisms, we calculate the probability density function (PDF) and the power spectral density (PSD) of the resistance fluctuations.  For this, we use all three data sets for the same experiment, such as the ones in Fig. \ref{fig:figure4}\textit{A}.  Fig. \ref{fig:figure4}\textit{B} shows the normalized PDFs in units of the rms fluctuation amplitude. The black data are the PDF of the background fluctuations collected in a device filled with just LB broth. These background fluctuations  are for the most part Gaussian, with  $\delta R_{bg} \approx 0.52~\rm k\Omega$. The blue data  obtained from \textit{E. coli} in nalidixic acid starts to deviate from a Gaussian and can be fitted by a stretched exponential function. The red data in ampicillin with the sharp peaks strongly deviates from a Gaussian. Fig. \ref{fig:figure4}\textit{C} shows the average  power spectral densities (PSDs) of these noise-like signals. The PSD of the signal in the bacteriostatic antibiotic (blue) is close to the PSD of the  noise without bacteria.  The added noise power due to the bactericidal effect (red) is at low frequencies in the range 0.01 Hz to 0.5 Hz, which is the high-frequency cut-off frequency in the measurement circuit. The $1/f$-like behavior of the PSD is probably due to the fact that the bursts take place on different time scales.  We further compare the rms value of the  fluctuations in ampicillin in the first half with that in the second half of the two-hour  measurement,  $ \delta R_{amp}^{(i)} \approx 0.5\delta R_{amp}^{(ii)}$, indicating that ampicillin exhibits time-dependent bactericidal effect \cite{levison2009pharmacokinetics,midolo1996bactericidal}.

\section*{Discussion}

This work describes an electrical approach that determines bacterial susceptibility to antibiotics in a microfluidic device. In the simplest interpretation, the approach depends on the blockage of (quasi direct current) ionic current by intact bacteria \cite{tsutsui2018identification}. Upon further reflection, however, deeper questions emerge on how  ionic current flows in pores and microchannels blocked by bacteria. Some of our microscope images suggest that, after cell lysis, the resistance tends to decrease even before the cell residue gets washed out by the liquid flow. We speculate that, once the cell wall and membrane lose their integrity, a bacterium may start to conduct ionic current  at a higher rate through the cell body. In the case of rapid cell  bursts, the residues are probably too small to significantly block the ionic current efficiently. 

At the current stage of development, the entire setup consists of  a basic Ohmmeter and a flow controller connected to the microfluidic device. Given that the approach  does not require a high-resolution microscope, it could eventually be developed into a small and robust point-of-care platform, potentially usable in resource-limited settings. A few technical improvements are still needed:   First, the sample loading process could be optimized. A higher applied pressure will allow the sample to flow faster and reduce the loading time \cite{baltekin2017antibiotic}. Currently, the pressure is limited by the bonding strength between PDMS and the substrate. A silicon-based device, while harder to fabricate, may solve this problem.  Second, the electrical measurement can  be multiplexed to increase the throughput,  reduce the test time, or  provide the susceptibility of  bacteria to multiple antibiotics in parallel. 

The fluctuations in the electrical signal due to antibiotic action are worth serious attention.  The fluctuations in the bacteriostatic antibiotic are close to those in LB broth, with the slight increase in the rms amplitude possibly being due to the movements of the  \textit{E. coli} \cite{kara2018microfluidic}. The more interesting question is the strong deviation of the fluctuations in the bactericidal antibiotic experiment from Gaussian statistics, approaching an exponential distribution. Apparently, the  short strong and discrete peaks generated by the cell bursts are responsible for the observed behavior. This is reminiscent of wall turbulence, where strong and rare turbulent wall bursts completely dominate the velocity fluctuations in a similar manner. A focused parametric experimental study and a first-principles theory is needed for a more complete biophysical picture. 

In the short term, our method is poised to have clinical relevance to UTI diagnosis and optimal treatment. While this work lays the foundation for an antibiotic susceptibility test, further translational studies are needed for a clinical test. Patients with UTIs and possible co-morbidities (e.g., diabetes, chronic renal disease) likely have complex urine matrices that may not be directly usable in our device. However, uncomplicated UTI in otherwise healthy adult women is one of the most common UTI syndromes in outpatient medicine, and the need for rapid susceptibility testing to improve empirical therapy is increasing with more community-based gram negative resistance \cite{talan2016fluoroquinolone}. Polymicrobial UTIs, in which multi-pathogens with heterogeneous antibiotics response co-exist, may require additional considerations.

\section{Materials \& methods}

\subsection{Microfluidic Device}

The microfluidic device consists of a PDMS microstructure (embedded with a two-layer microfluidic channel) that is permanently bonded with a glass substrate, which has metallic electrodes on it. We use standard soft lithography to fabricate the device. Details of the device fabrication process are described in SI Appendix, Supplemental Materials and Methods.

\subsection{Bacterial Strains, Growth Media and Antimicrobial Preparations}

In this study, motile \textit{E. coli} (ATCC  25922), \textit{K. pneumoniae} (ATCC 13883),   \textit{S. saprophyticus} (ATCC  15305) were purchased from American Type Culture Collection (ATCC, Manassas, VA), and non-motile \textit{E. coli} (JW 1908-1)  was obtained from \textit{E. coli} Genetic Stock Center (New Haven, CT). We used either Luria-Bertani (LB) broth (Sigma-Aldrich, St. Louis, MO) or human urine (Lee Biosolutions, Maryland Heights, MO) as growth media depending on the experiment.  The stock solutions of ampicillin and nalidixic acid were prepared using the methods provided by the supplier (Alfa Aesar, Ward Hill, MA). Details are provided in SI Appendix, Supplemental Materials and Methods.

\subsection{Electrical Measurements and Data Acquisition}

A lock-in amplifier (SR 830 DSP, Stanford Research Systems, Sunnyvale, CA) is used to measure the resistances of the microchannels. The reference frequency and time constant are 10 Hz and 300 ms (bandwidth of $\sim 0.53$ Hz), respectively. The output signals from the lock-in amplifier were recorded using a data acquisition card (NI 6221, National Instruments, Austin, TX) through a LabVIEW (National Instruments, Austin, TX) Virtual Instrument (VI) interface. The sampling rate for data collection is 6 Hz. The experimental data are analyzed using Origin (MicroCal Software, Northampton, MA) and MATLAB (Mathworks, Natick, MA). Details of electrical measurements are given in SI Appendix, Supplemental Materials and Methods. 

\subsection{Image Collection}

Images of the bacterial cells in the microchannels were obtained in an Axio observer inverted microscope (Carl Zeiss, Jena, Germany)  using a 63$\times$ objective, an AxioCam 503 mono camera (Carl Zeiss, Oberkochen, Germany), and ZEN image acquisition software (Carl Zeiss, Jena, Germany).

\subsection{Data Availability}

 All data and procedures are included in the manuscript, SI Appendix, and Movies S1 and S2.


\section{Acknowledgements}

This work was supported by the NIH (1R21AI133264-01A1 and 1R03AI126168-01). The authors thank Joan O'Connor for  assistance with sample preparation and resazurin-based broth microdilution ASTs and Deborah J. Stearns-Kurosawa and Victor Yakhot for discussions.

\bibliography{Main}

\end{document}


\title{Supplementary Information for\\ ``All-electrical monitoring of bacterial antibiotic susceptibility in a microfluidic device''}

\author{{ Yichao Yang$^1$, Kalpana Gupta$^2$, and Kamil L. Ekinci$^1$}\\
{\small \em $^1$Department of Mechanical Engineering, Boston University, Boston, Massachusetts 02215, USA\\
$^2$Department of Medicine, Boston University School of Medicine, Boston, Massachusetts 02118, USA}}

\date{\today}
\maketitle

\section{Supplemental Materials and Methods}

\section{Device Design and Fabrication}

In designing the device, we used a first-pass optimization to determine the number $k$ of parallel microchannels. The throughput increases (and the  loading time decreases) with increasing $k$. However, the available signal from a microchannel is divided between $k$ parallel resistors (see the resistance change calculation due to a single bacterium below). The use of $k=10$ parallel microchannels allowed us to achieve a loading time $\lesssim 30$ min and to  comfortably observe resistance changes due to single cells.

Molds for the two-layer microfluidic channel are fabricated by patterning  SU-8 photoresist (Microchem, Newton, MA) onto a 4-inch silicon wafer. After mixing pre-polymer with cross-linker (Sylgard 184, Dow Corning, Midland, MI) at a 9:1 ratio, the mixture is degassed in a vacuum desiccator for 30 minutes. Next, the  bubble free PDMS mixture is slowly poured onto the SU-8 mold and cured in a $90^{\circ}$C oven for 1 hour. The slab of PDMS with the embedded two-layer microfluidic channel structure is carefully peeled off from the master. Inlet and outlet ports (0.75 mm diameter) are mechanically punched into the PDMS using a biopsy punch. The PDMS structure and a glass slide with pre-defined metallic electrodes are sterilized and bonded through oxygen plasma treatment. To fabricate the chromium (Cr) and gold (Au) electrodes onto the glass slide, we use electron beam evaporation. The electrodes are fabricated by evaporating a 90-nm-thick Au layer on top of a 60-nm-thick Cr adhesion layer.

\section{Bacteria Culturing}

First, lyophilized bacteria are re-solubilized and mixed gently with  1 mL of LB broth, and  the solution is transferred into 5 mL of LB broth for each bacterial strain. Next, the bacteria  are grown in a shaking incubator at $37^{\circ}$C and 100 rpm for 24 hours. After 24 hours, the turbid bacterial suspension is centrifuged for 6 minutes at 6000 rpm, and the bacteria pellet is re-suspended in 5 mL of fresh LB broth. Finally,  frozen stocks are prepared by dissolving highly purified glycerol (MP Biomedicals, Solon, OH) at 20\% v/v in PBS (Lonza BioWhittaker, Walkersville, MD), mixing with bacterial suspension at 1:1 ratio, collecting into 200 $\mu$L aliquots and storing at $-80^{\circ}$C. On the day prior to the experiment, a frozen stock is thawed, of which 150 $\mu$L is  transferred into 8 mL of fresh LB broth;  10 $\mu$L of the bacterial culture is streaked on a LB agar (Becton Dickinson, Sparks Glencoe, MD) plate and grown overnight to check the purity of the bacterial culture. Bacteria are cultured overnight at 37$^{\circ}$C in a shaking incubator at 100 rpm. On the day of the experiment, the bacteria culture is diluted to the desired concentration. We measure the optical density of the culture at a wavelength of 600 nm (OD$_{600}$) using a spectrophotometer (V-1200, VWR, Radnor, PA). An OD$_{600}$ of 0.1 corresponds to a bacterial cell density of $ 2\times 10^{7}$ CFU/mL, which is  periodically confirmed through serial dilution plating on LB agar plates.

\section{Resazurin-Based Broth Microdilution AST}

In order to compare our method with standard methods, the susceptibility of \textit{E. coli}, \textit{K. pneumoniae}, and \textit{S. saprophyticus} are determined using resazurin-based broth miltidilution AST standardized by the Clinical and Laboratory Standard Institute (CLSI). To ensure consistency,  \textit{E. coli} (ATCC 25922), for which the MICs of both ampicilin and nalidixic acid are 4 mg/L \cite{andrews2001determination},  is used as a reference strain for all the resazurin-based microdilution tests.  First, ampicillin and nalidixic acid are diluted from stock solutions in LB broth. Column 12 of the 96-well plate is used as  growth control (no antibiotics); column 11 is used as  sterility control (no bacteria); and columns 1-10 are filled with solutions with decreasing antibiotic concentrations, which are prepared by using the two-fold serial dilution method. Next, the bacterial cultures are prepared separately at $37^{\circ}$C in 8 mL of LB broth. After adjusting their OD$_{600}$  to 0.1, the solution is further diluted by a factor of 20 with LB broth. Then, 100 $\mu$L of bacteria solution is added to each well of columns 1 to 10, and column 12. The final bacterial concentration in each well is  5$\times 10^{5}$ CFU/mL. The bacterial suspensions are used within 30 min after their optical density are adjusted to avoid changes of cell numbers \cite{wiegand2008agar}. Each concentration is replicated in three wells in each plate. After incubating the plate at 37$^{\circ}$C in a shaking incubator at 100 rpm for 16-20 hours, 60 $\mu$L 0.015\% solution of resazurin (ACROS Organics, New Jersey, USA) in tissue culture grade water is added to each well and further incubated at 37$^{\circ}$C for another 4 hours. The plate results are read by visual inspection of the wells. Dark blue/purple indicates that bacteria are not viable, and pink  indicates that bacteria are still viable. If the growth control shows dark blue/purple or the sterility control indicates contamination, the plate is discarded. The MICs determined using the resazurin-based broth miltidilution AST are summarized in Table \ref{tab:tableS1}.

\section{Experimental Protocol for Electrical Measurements}

We measure the growth of \textit{E. coli}, \textit{K. pneumoniae}, and \textit{S. saprophyticus} in LB broth with and without antibiotics. After adjusting an overnight bacterial culture to an OD$_{600}$ of 0.1, the culture is diluted 1:20 into 5 mL LB broth and, depending on the experiment, mixed with antibiotics in equal volume. This results in a final bacterial cell density of $5\times 10^{5}$ CFU/mL. The mixture is transferred into a sterile 15-mL Falcon tube that is used as a sample reservoir. A fluorinated ethylene propylene (FEP) tube (Cole-Parmer, Vernon Hills, IL) is used to connect the sample reservoir to the microfluidic device inlet. During sample loading, the inlet is pressurized at $\Delta p \sim 10$ kPa above the outlet. The number of trapped bacteria is typically not uniform across microchannels. After approximately tens of bacteria are trapped in the microchanels, voltage drop across the microchannels is measured to quantify the bacterial growth. During the measurements, the pressure difference between inlet and outlet is maintained at  $\Delta p \sim 0.5$ kPa.  The pressure during loading is controlled using a pressure controller (OB1-Mk3, Elveflow, Paris, France). To ensure a stable temperature of the microfluidic device during an experiment, a PeCon 2000–2 Temp Controller (PeCon GmbH, Erbach, Germany) is used. To show that our microfluidic device can be used to determine MICs for antibiotics in human urine samples, we measure \textit{K. pneumoniae} in nalidixic acid  and \textit{E. coli} (non-motile) in ampicillin. Bacteria concentration is adjusted to an OD$_{600}$ of 0.1, and the bacteria solution is diluted 1:20 in 5 mL of human urine sample. Subsequently, the bacteria-spiked urine samples are mixed with LB broth and  nalidixic acid (0, 4, 8, 16, and 32 mg/L) or ampicillin (0, 2, 4, 8, and 16 mg/L) in equal volume prior to loading into a microfluidic device.

\section{Electrical Measurements}

\begin{figure}[t]
\centering
\includegraphics[width=5 in]{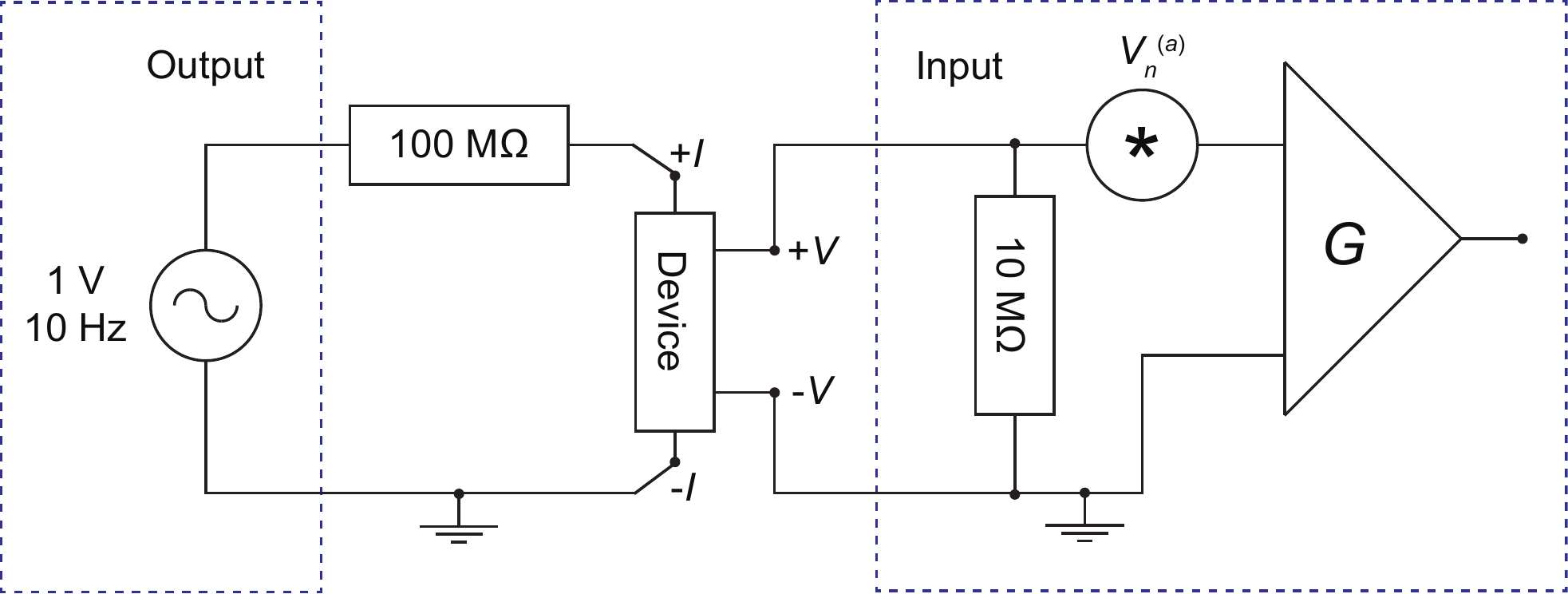}
\caption{Equivalent-electrical circuit for the measurement. The dashed boxes represent the lock-in amplifier; $V_{n}^{(a)}$ is the input noise voltage and $G$ is the gain of the lock-in amplifier.}
\label{fig:figureS1}
\end{figure}

Fig. \ref{fig:figureS1} shows the simplified equivalent circuit model of the electrical measurement. The lock-in amplifier oscillator output $V_{s}$ (rms amplitude of 1 V and reference frequency of $f_{r}=10~ \rm Hz$) is connected to a resistor $R_{s}=100 ~\rm M\Omega$ to create a current source, which drives a current (several nA)   through the device and the input circuit of the lock-in. The input resistance of the lock-in amplifier is $10 ~\rm M\Omega$. We use a  four-wire  measurement to measure the resistance of the device. At $f_{r}=10~ \rm Hz$, the four-wire electrical impedance of the device is dominated by its resistance; typical device impedance at the start of each experiment is $\approx 3-0.2{i} ~\rm M\Omega$, corresponding to a phase angle of $-4^{\rm o}$. The resistance value and the phase both drift over the course of two hours. The  drift in the resistance is  $\lesssim 1\%$ and the impedance phase angle is $\pm 1^{\rm o}$.  We estimate  each contact impedance to be $\approx 70-{700}{i} ~\rm k\Omega$ at 10 Hz through two-wire measurements. We use a 300 ms time constant on the lock-in amplifier and digitally sample the data from the lock-in at a rate of 6 Hz. When we focus on the long-term behavior of the resistance (e.g., growth or antibiotic susceptibility measurements over two hours), we further integrate (average) the data numerically over one-minute intervals. When we focus on the short-time fluctuations (i.e., Fig. 4 in main text), we high-pass filter the data using a cut-off frequency of $0.01$ Hz.

The time-dependent resistance of the device can be expressed as $R(t) = {R_{em}} + \Delta R(t)$, where $R_{em}$ is the initial resistance of the microchannels with pure LB broth and $\Delta R(t)$ is the resistance change induced by bacteria in the microchannels. We  can estimate  the minimum detectable $\Delta R$ from noise analysis.  In the experiments, the equivalent noise bandwidth at a time constant of 300 ms is $\Delta f \approx 0.31~ \rm Hz$  (time constant  300 $\rm ms$ and filter roll-of 18 $\rm dB/oct$). To determine the experimental noise floor, we perform a noise measurement using a $3.2~\rm M\Omega$ source resistor. We obtain a total noise of $ \sim 600~\rm nV/ {Hz^{1/2}}$. This value is slightly larger than the theoretical value of $\sim 300$ $\rm nV/Hz^{1/2}$, obtained from combining the Johnson noise of a $3.2~\rm M\Omega$ resistor (230 $\rm nV/Hz^{1/2}$) with the input noise, $V_{n}^{(a)}$, of  the lock-in at 10 Hz  ($V_{n}^{(a)}\lesssim 200$ $\rm nV/Hz^{1/2}$). We use a low-sensitivity setting on the lock-in to be able to track the large changes in the device resistance during bacteria growth. The minimum detectable resistance change or the resistance noise can be estimated from a simple circuit analysis. Here, we assume that the minimum detectable resistance change (under the imposed current of 10 nA across the device and the lock-in input) results in a voltage equal to the noise voltage. This provides  $\approx 200~\rm \Omega$, which is close to the resistance fluctuations (noise) observed in LB broth. 

In order to quantify the effect of the long-term electrical drifts on the sensitivity, we have performed a set of experiments using devices clogged with 1-$\mu \rm m$-diameter polystyrene (PS) microspheres.  This is similar to clogging the microchannels with bacteria but, since the PS microspheres do not change in size over time, we are able to extract the electrical drift under conditions  comparable to  bacteria  experiments. In particular, we clog the devices to resistance values $R(0)$ (or $\Delta R(0)$) close to those in bacteria experiments, indicating similar flow rates and initial conditions. Three baseline resistance drifts measured over the course of 2 hours   are shown in Fig. \ref{fig:figureS2}\textit{A}. Since the drift appears linear, we fit it as $\Delta R (t) -\Delta R(0) \approx -0.26t$ (in units of $\rm k\Omega$ when $t$ is in minutes). In an effort to quantify the drift effect on the antibiotic susceptibility tests, we have recalculated the drift-corrected growth rates  (dashed lines in Fig. \ref{fig:figureS2}\textit{B-F}).  Here, the solid lines are the results from Fig. 3  in the main text. In the recalculation, we first subtracted the drift from each data trace and then computed the growth rate. Our conclusion, after comparing the growth rates of corrected and raw data in Table \ref{tab:tableS2}, is that drift can safely be neglected at this stage of development.

\section{Resistance Change Per Added Bacterium}

We show the resistance change $\Delta R$ as a function of the number $n$ of bacteria  in the microchannels for \textit{K. pneumoniae}, \textit{E. coli}, and \textit{S. saprophyticus} in Fig. \ref{fig:figureS3}. Data shown in each figure are from three independent experiments. Red dashed lines are the linear fits to the data. We obtain  $\Delta R_{1}^{(KP)} \approx 2.5 \pm 0.3 ~\rm k\Omega$  for  \textit{K. pneumoniae} (Fig. \ref{fig:figureS3}\textit{A}), $\Delta R_{1}^{(EC)} \approx 3.7 \pm 0.3 ~\rm k\Omega$ for  \textit{E. coli}  (Fig. \ref{fig:figureS3}\textit{B}), and $\Delta R_{1}^{(SS)} \approx 3.5 \pm 1.1 ~\rm k\Omega$ per \textit{S. saprophyticus}  (Fig. \ref{fig:figureS3}\textit{C}). The larger error in \textit{S. saprophyticus}  originates from the fact that it is more challenging to count single cells from microscope images and cells tend to cluster more.

The measured $R_{em}$ is the equivalent resistance of  ten parallel microchannels at the center of the microfluidic device: $R_{em} = {1\over 10} {R_{em}^{(s)}}$, where $R_{em}^{(s)}$ is the single microchannel resistance, $R_{em}^{(s)}=\rho{\frac{l}{A}}$, with $\rho$ being the electrical resistivity of the liquid media (e.g., LB broth) filling the microchannel, $l$ and $A$ being respectively the length and cross-sectional area of the single microchannel.  We assume that the electrical resistance of bacteria is large compared to the media. Thus, the resistance of a single microchannel with one trapped bacterium can be estimated as $R_{em}^{(s)}+ \Delta R_{1}^{(s)} \approx \rho \left[ \frac{l-l_B}{A}+\frac{l_B}{A-A_B}\right]$, where $l_B$ and $A_B$ are the length and cross-sectional area of a bacterium, respectively. Here, $R_{em}\approx 3 ~\rm M\Omega$ in LB broth, which, using the nominal channel dimensions, gives $\rho\approx 1.2~\rm \Omega \cdot m$.  \textit{K. pneumoniae} is  rod-shaped, with  $l_B = 2 ~\mu {\rm m}$ and $A_B = 0.8 ~\mu \rm m^2$ \cite{berg2008coli}. Using these numbers, we obtain $\Delta R_{1}^{(s)} \approx 150 ~\rm k\Omega$. Calculating the equivalent resistance, we find the total resistance change  per bacterium  becomes $\Delta R_{1} \approx \frac{\Delta R_{1}^{(s)}}{100} \approx 1.5~\rm k \Omega$. Note that the resistance change per bacterium very much depends on the size of the bacterium and how the bacterium blocks the microchannel during growth. It is thus different for \textit{S. saprophyticus} and \textit{E. coli}.

\section{Bacteria Accumulation or Growth Outside of the Microchannels} 

 We occasionally observe  bacteria accumulation outside of the microchannels or growth outward. Fig. \ref{fig:figureS4} shows two different non-ideal ways  bacteria accumulate in the device. The linear dimensions of the  regions immediately upstream and downstream from the microchannels are $l\times w\times h \approx 75\times 80 \times 2 ~\mu \rm m^3$. The microscope images in Fig. \ref{fig:figureS4} show a portion of this region in addition to the central microchannels. Fig. \ref{fig:figureS4}\textit{A} shows that bacteria (\textit{E. coli})  can get immobilized in the inlet region; in addition, any bacteria that escapes through the nanoconstriction can proliferate in the outlet region. Fig. \ref{fig:figureS4}\textit{B} shows that bacteria (\textit{S. saprophyticus}) can get stuck at the entry region of the microchannels, blocking further bacteria trapping in the microchannels. When bacteria are trapped in these bigger channels ($l\times w\times h \approx 75\times 80 \times 2 ~\mu \rm m^3$), the resistance change per added bacterium no longer follows the ideal case discussed in the main text. In fact from geometry,  the resistance change per bacterium is $\sim {1\over 20} \Delta R_{1}$, where $\Delta R_{1}\approx 1.5~\rm k \Omega$ is the  resistance change per added bacterium into  one of the ten smaller microchannels, as discussed above. For bacteria accumulating outside of this 2-$\mu$m-high region, the resistance change is even smaller. Thus, our measured resistance signals mainly come from the ten smaller microchannels. We note that  bacteria trapped outside the microchannels also grow (or die). Thus, their signals are coherently added to the signals developing in the microchannels. Finally, if the experiment continues for a long time, bacteria, especially motile strains, tend to escape more readily and/or grow outward after filling the microchannels.  

\section{Estimation of Bacterial Doubling Time}

We have estimated the doubling time $t_{d}$ of bacteria using resistance change data during growth. As discussed in the main text, $\frac{{\Delta R(t)}}{{\Delta R(0)}} \approx \frac{{n(t)}}{{n(0)}} = {e^{\frac{{\ln 2}}{{{t_d}}}t}}$. Thus, $t_d$ can be obtained by  a linear fit to the natural logarithm of ${{\Delta R (t)}\over{\Delta R (0)}}$. Fig. \ref{fig:figureS5} shows a number of fits (dashed lines) to the experimental resistance data (solid lines) for \textit{E. coli}, \textit{K. pneumoniae}, and \textit{S. saprophyticus}. Each  curve is from an independent experiment. The  $t_d$ and $R^2$ values are as indicated in the figure. 

\section{Data from All Measurements}

Fig. \ref{fig:figureS6} shows the resistance change, $\Delta R(t)-\Delta R(0)=R(t)-R(0)$,  measured over the course of 2 hours after sample loading in  all our antibiotic susceptibility tests and growth experiments. For each data plot, the right $y$ axis shows the change in the number of bacteria in the device, $\Delta n(t)$, which is  estimated from $\Delta n(t)={{\Delta R(t)-\Delta R(0)} \over \Delta R_{1}}$ with $\Delta R_1$ being the calibration value from Fig. \ref{fig:figureS3}. The initial resistances $R(0)$ measured at the start of each electrical measurement are shown in Table \ref{tab:tableS3}. The approximate number of trapped bacteria in the microchannels at the start of each electrical measurement from microscope images are listed in Table \ref{tab:tableS4}. 

\section{Metric for Assessing Antibiotic Susceptibility} 

\begin{figure}[t]
\centering
\includegraphics[width=6.75 in]{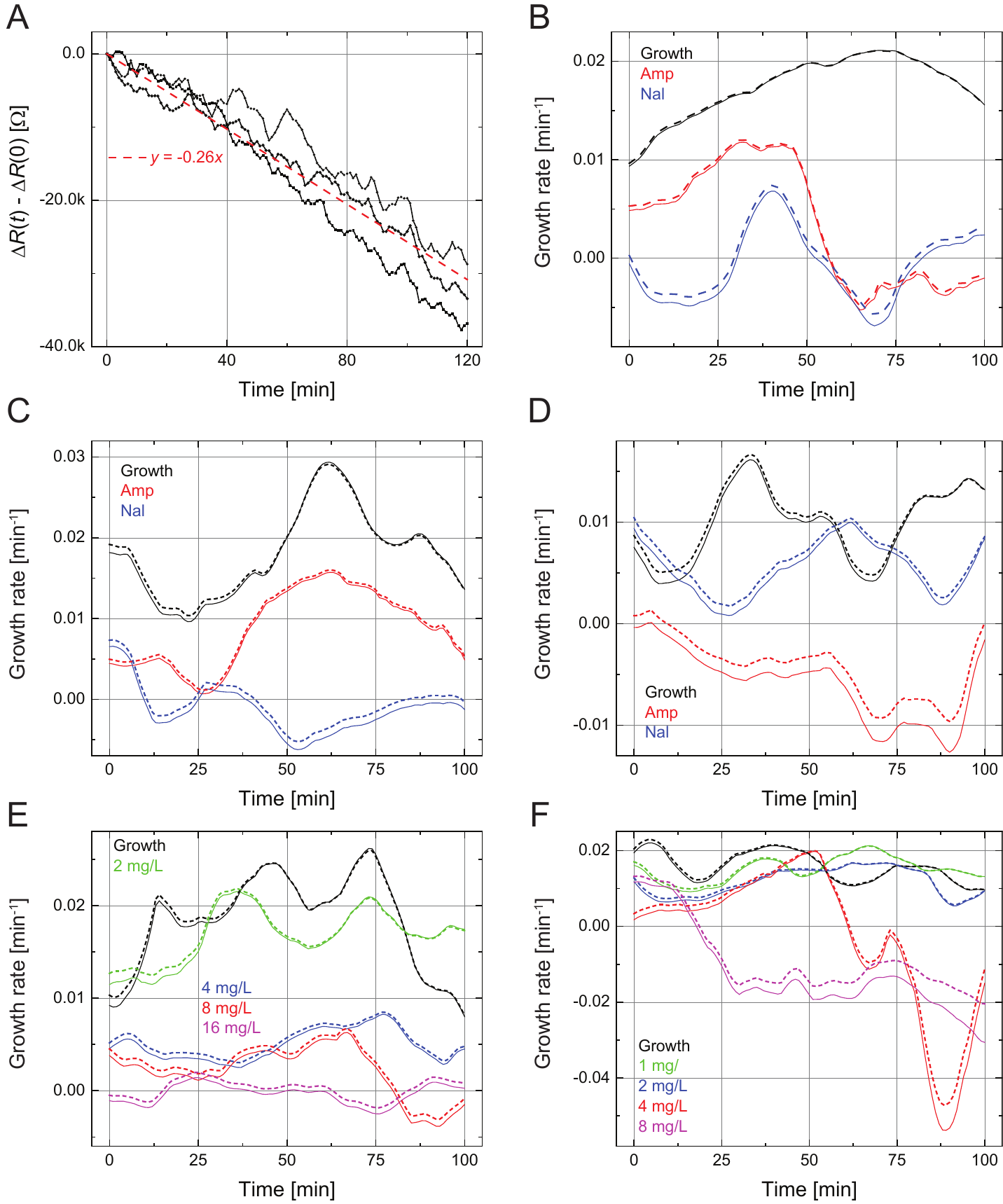}
\caption{(\textit{A}) Baseline drifts as a function of time in three independent experiments; the red dashed line shows the linear fit. (\textit{B-F}) Growth rates from raw   (solid lines) and drift-corrected resistance data (dashed lines). Solid lines are reproduced from Fig. 3 in the main text.}

\label{fig:figureS2}
\end{figure}

\begin{figure}[t]
\centering
\includegraphics[width=6.75 in]{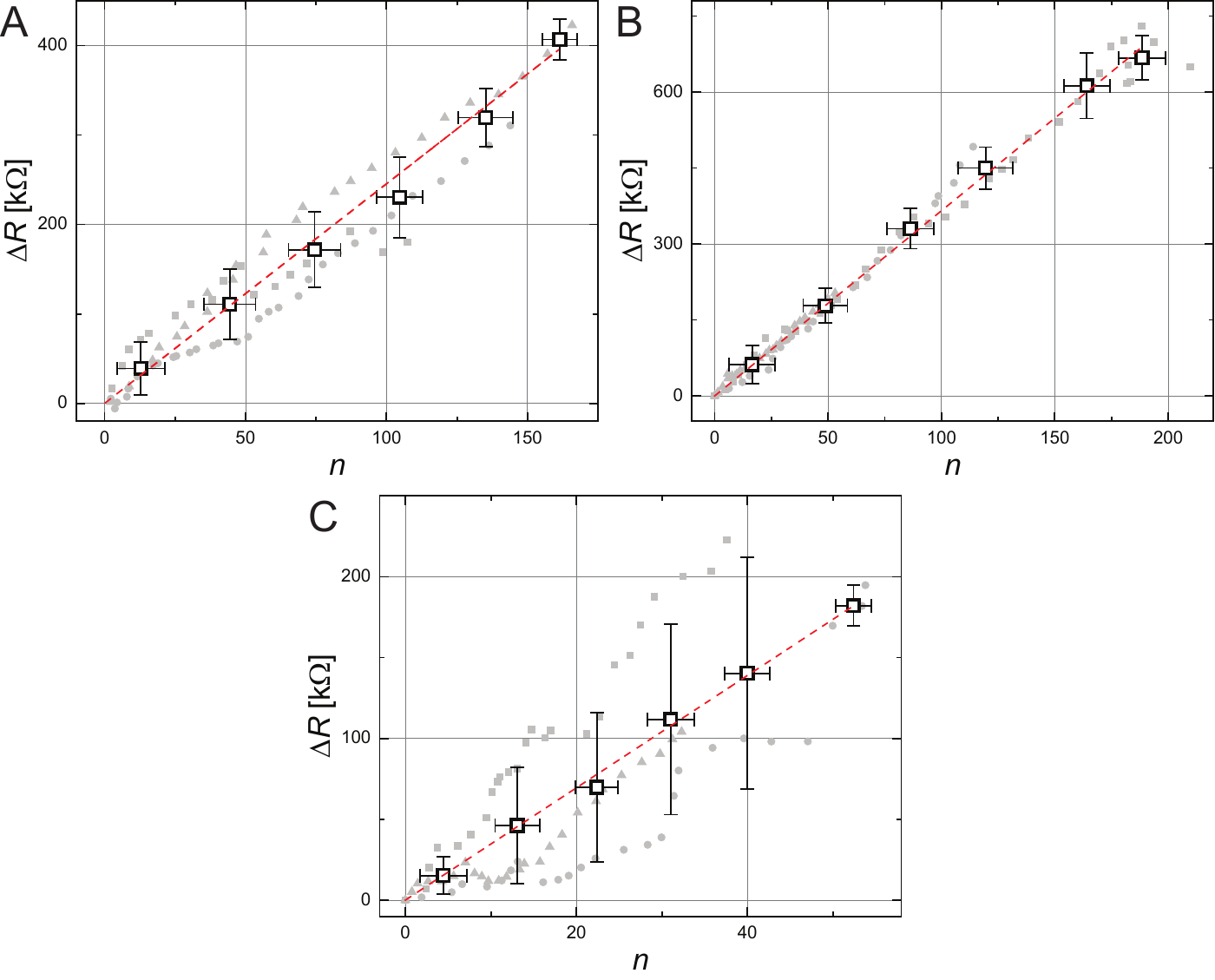}
\caption{Resistance change $\Delta R$ as a function of the number $n$ of bacteria in the microchannels from three independent experiments. (\textit{A}) \textit{K. pneumoniae}. (\textit{B}) \textit{E. coli}.  (\textit{C}) \textit{S. saprophyticus}. The red dashed line in each figure shows the linear fit.}
\label{fig:figureS3}
\end{figure}

\begin{figure}[t]
\centering
\includegraphics[width=6.75 in]{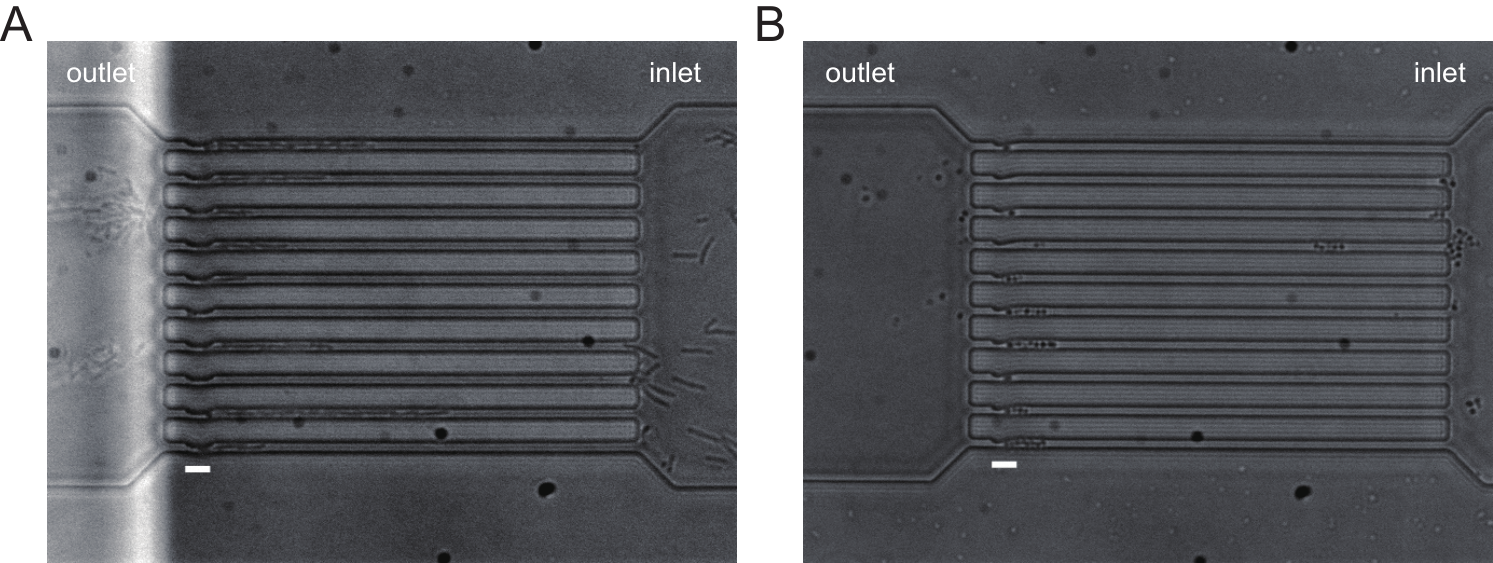}
\caption{Microscope snapshots showing bacteria accumulation outside the microchannels. (\textit{A}) \textit{E. coli} (ATCC 25922) growing in LB broth in nalidixic acid (20 mg/L). The bacteria that escaped through the nanoconstriction have proliferated in the outlet region.  (\textit{B}) \textit{S. saprophyticus} (ATCC 15305) growing in LB broth in ampicillin (10 mg/L). Some cells have accumulated at the entry regions of the microchannels. The  scale bars are 5 $\mu$m.}
\label{fig:figureS4}
\end{figure}

\begin{figure}[t]
\centering
\includegraphics[width=6.75 in]{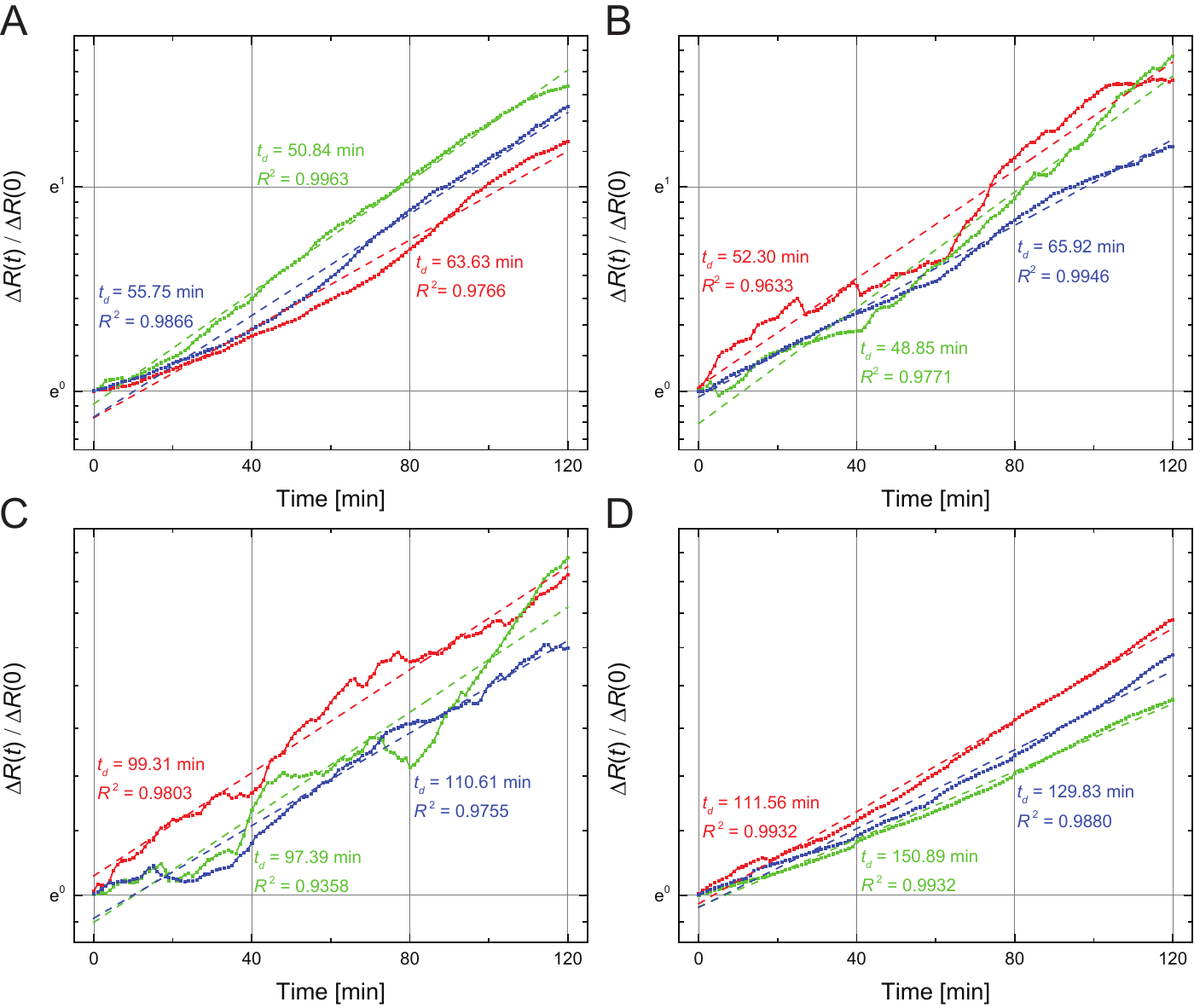}
\caption{Linear fits  to the natural logarithm of $\Delta R(t)\over \Delta R(0)$ in order to determine the bacterial doubling time in the microchannels. (\textit{A}) \textit{E. coli} (ATCC 25922) at $37~^{\circ}$C. (\textit{B}) \textit{K. pneumoniae} (ATCC 13883) at $37~^{\circ}$C. (\textit{C}) \textit{S. saprophyticus} (ATCC 15305) at $37~^{\circ}$C.  (\textit{D}) \textit{E. coli} (ATCC 25922) at $23~^{\circ}$C. Solid lines show data from  independent experiments; the dashed lines show the linear fits.}
\label{fig:figureS5}
\end{figure}

\begin{figure}[t]
\centering
\includegraphics[width=6.75 in]{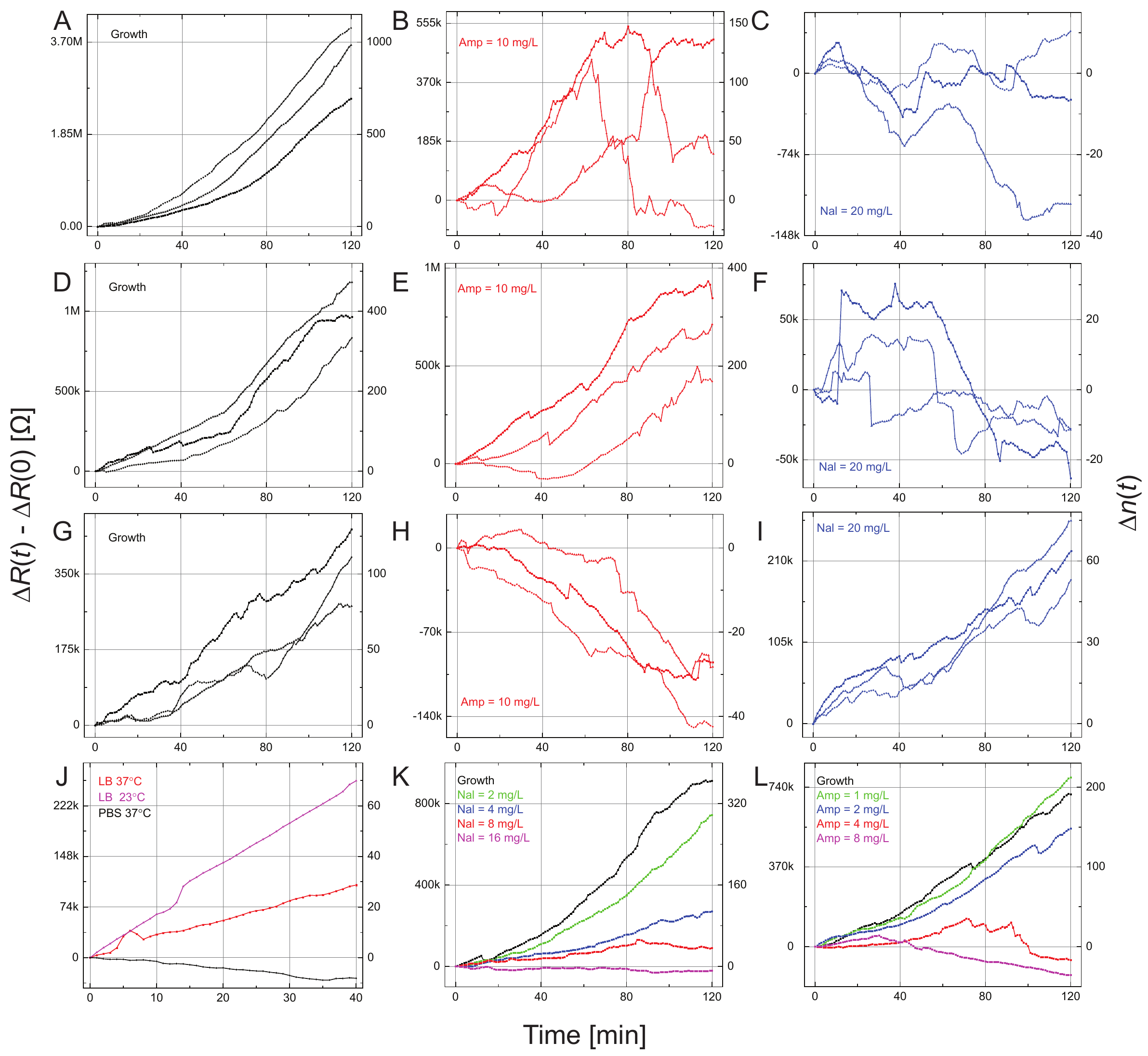}
\caption{Measured resistance change $\Delta R(t)-\Delta R(0) $ and change  in the number of bacteria, $\Delta n(t)$, in the microchannels after sample loading as a function of time.  Measurements on \textit{E. coli} (ATCC 25922) (\textit{A-C}), \textit{K. pneumoniae} (ATCC 13883) (\textit{D-F}), \textit{S. saprophyticus} (ATCC 15305) (\textit{G-I}). (\textit{J}) Measurements on \textit{E. coli} (ATCC 25922)  in PBS at $37^{\circ}$C and LB broth at $23^{\circ}$C and $37^{\circ}$C. (\textit{K}) Measurements on \textit{K. pneumoniae} (ATCC 13883)  growing in urine with nalidixic acid at different concentrations. (\textit{L}) Measurement on \textit{E. coli} (JW 1908-1) growing in urine with ampicillin at different concentrations. Each curve represents one independent experiment.}
\label{fig:figureS6}
\end{figure}

A simple metric for aility can be obtained from the time derivative of the resistance data,  $\frac{d}{{dt}}\ln \left[ {\frac{{\Delta R(t)}}{{\Delta R(0)}}} \right]$. Given that ${{\Delta R (t)}\over{\Delta R (0)}} \approx {{n (t)}\over{n(0)}}$  and  $n(t) \approx n(0){e^{rt}}$,  $\frac{d}{{dt}}\ln \left[ {\frac{{\Delta R(t)}}{{\Delta R(0)}}} \right] \approx r$. If $r>0$, the population grows; if $r\le0$, the population does \textit{not} grow. Since $r$ itself is a function of time, especially for antibiotics acting with some delay, it may be more appropriate to consider $r$ averaged over roughly the second half of the experiment. Thus, we  calculate  $\bar r$ averaged over the last 40 mins of available data, which corresponds to the last 60 mins of the resistance measurement due to  the 20-min time window of the derivative. Table \ref{tab:tableS2} shows  $\bar r$ values in all experiments calculated from  raw data as well as  drift-corrected data  (Fig. \ref{fig:figureS2}\textit{B-F}). This metric provides conclusions consistent with standard AST results. We note, however, that $\bar r \le 0$ may be too restrictive a condition for susceptibility, especially for a clinical application. More data and error analysis may allow us to relax this condition to $\bar r \le \varepsilon$, where  $\varepsilon>0$.

\clearpage

\clearpage

\clearpage

\clearpage

\clearpage

\clearpage

\clearpage

\begin{table*}[t]
\caption{Summary of MICs of ampicillin and nalidixic acid for \textit{E. coli}, \textit{K. pnuemoniae} and \textit{S. saprophyticus} obtained by  resazurin-based broth multidilution AST. Results obtained in urine  are shown in parentheses.}
\begin{ruledtabular}
\begin{tabular}{ccccc}
Bacteria  & Ampicillin (mg/L) & Nalidixic acid (mg/L)\\
\hline
\textit{E. coli} (ATCC 25922) & 8 & 4-8\\
\textit{K. pnuemoniae} (ATCC 13883) & $>$ 128 & 16 (8)\\
\textit{S. saprophyticus} (ATCC 15305) & $<$ 0.25 & $>$ 128\\
\textit{E. coli} (JW 1908-1) & (4) &-\\
\end{tabular}
\end{ruledtabular}
\label{tab:tableS1}
\end{table*}

\clearpage

\begin{table*}[t]
\caption{Average growth rates $\bar r$ for all experiments without and with drift correction. Where available, both the average values and the results of individual experiments (in parentheses) are tabulated.}
\begin{ruledtabular}
\begin{tabular}{ccccc}
 Bacteria & Antibiotic & Expectation & $\bar r$ (min$^{-1}$) & Drift-corr. $\bar r$ (min$^{-1}$)\\
\hline

\textit{E. coli} & Growth (LB)   & - & 0.019 (0.021, 0.018, 0.019) & 0.019 (0.021, 0.018, 0.019)\\
\textit{E. coli} & Ampicillin, 10 mg/L & Susceptible & -0.0037 (-0.0005, -0.0099, -0.0006) & -0.0027 (0.0003, -0.0093, 0.0008) \\
\textit{E. coli} & Nalidixic acid, 20 mg/L & Susceptible & -0.0017 (-0.0011, 0.0012, -0.0052) & -0.0008 (-0.0005, 0.0023, -0.0043) \\

\textit{K. pnuemoniae} & Growth  (LB) & - & 0.021 (0.022, 0.026, 0.015) & 0.021 (0.022, 0.026, 0.015)\\
\textit{K. pnuemoniae} & Ampicillin, 10 mg/L & Resistant & 0.0114 (0.0096, 0.0176, 0.0070)  & 0.0118 (0.0099, 0.0184, 0.0071) \\
\textit{K. pnuemoniae} & Nalidixic acid, 20 mg/L & Susceptible & -0.0019 (-0.0044, 0.0013, -0.0027) & -0.0010 (-0.0034, 0.0019, -0.0014)\\

\textit{S. saprophyticus} & Growth (LB) & - & 0.010 (0.007, 0.015, 0.009) & 0.010 (0.007, 0.015, 0.009)\\
\textit{S. saprophyticus} & Ampicillin, 10 mg/L & Susceptible & -0.0096 (-0.0058, -0.0068, -0.0161) & -0.0071 (-0.0039, -0.0051, -0.0124) \\
\textit{S. saprophyticus}  & Nalidixic acid, 20 mg/L & Resistant & 0.0059 (0.0052, 0.0055, 0.0069) & 0.0064 (0.0057, 0.0062, 0.0072) \\

\textit{K. pnuemoniae} & Growth (urine) & - & 0.018 & 0.018\\
\textit{K. pnuemoniae} & Nalidixic acid, 2 mg/L & Resistant & 0.018 & 0.018 \\
\textit{K. pnuemoniae} & Nalidixic acid, 4 mg/L & Resistant & 0.0059 & 0.0062 \\
\textit{K. pnuemoniae} & Nalidixic acid, 8 mg/L & Resistant & 0.0006 & 0.0012 \\
\textit{K. pnuemoniae} & Nalidixic acid, 16 mg/L & Susceptible & -0.0008 & -0.0002\\

\textit{E. coli} & Growth (urine) & - & 0.013 & 0.013\\
\textit{E. coli} & Ampicillin, 1 mg/L & Resistant & 0.017 & 0.017 \\
\textit{E. coli}  & Ampicillin, 2 mg/L & Resistant & 0.0125  & 0.0126 \\
\textit{E. coli} & Ampicillin, 4 mg/L & Susceptible & -0.0232 & -0.0197 \\
\textit{E. coli} & Ampicillin, 8 mg/L & Susceptible & -0.0193 & -0.0136 \\

\end{tabular}
\end{ruledtabular}
\label{tab:tableS2}
\end{table*} 

\clearpage

\begin{table*}
\caption{Initial resistances $R(0)$ for each electrical measurement. Experiments in urine  are shown in parentheses.}

\begin{ruledtabular}
\begin{tabular}{ccccc}
Bacteria & Growth ($\rm M\Omega$) & Ampicillin ($\rm M\Omega$) & Nalidixic acid ($\rm M\Omega$)\\

\hline

\textit{E. coli} ATCC 25922 & 4.02, 4.10, 4.15 & 3.47, 3.95, 3.99 & 3.59, 3.24, 3.59\\
\textit{K. pnuemoniae} ATCC 13883 & 3.22, 3.15, 3.46 & 3.65, 3.47, 3.91
 & 3.41, 3.55, 3.28 (3.24, 3.21, 3.52, 3.45, 3.60)\\
\textit{S. saprophyticus} ATCC 15305 & 3.28, 3.22, 3.24 & 3.29, 3.37, 3.19 & 3.28, 3.16, 3.43\\
\textit{E. coli} JW 1908-1 & - & (3.23, 3.27, 3.26, 3.18, 3.16) & -\\

\end{tabular}
\end{ruledtabular}
\label{tab:tableS3}
\end{table*}

\clearpage

\begin{table*}
\caption{Rough number of  bacteria trapped in the microchannel region in each experiment as determined from microscope images. Experiments in urine are shown in parentheses.}

\begin{ruledtabular}
\begin{tabular}{ccccc}
Bacteria & Growth & Ampicillin & Nalidixic acid\\

\hline

\textit{E. coli} (ATCC 25922) & 90, 95, 95 & 70, 95, 95 & 75, 50, 75\\
\textit{K. pnuemoniae} (ATCC 13883) & 60, 50, 85 & 85, 65, 95
 & 60, 70, 50 (60, 60, 70, 65, 80)\\
\textit{S. saprophyticus} (ATCC 15305) & 60, 60, 50 & 45, 60, 45& 40, 50, 65\\
\textit{E. coli} (JW 1908-1) & - & (60, 55, 40, 30, 35) & -\\

\end{tabular}
\end{ruledtabular}
\label{tab:tableS4}
\end{table*}

\clearpage

\section{Supplemental Movies} 

Movies S1. \textit{E.coli} (ATCC 25922) growth with no drug in the microchannels in LB broth at $37^{\circ}$C. (Left) Time-lapse imaging showing that the cells are immobilized and growing in the microchannels. Scale bar, 5 $\mu$m. (Right) The normalized electrical resistance change ${{\Delta R(t)} \over {\Delta R(0)}}$ of the microchannels as a function of time. Each second in the video corresponds to $\sim3$ min in the experiment.

\bigbreak

Movies S2. \textit{E. coli} (ATCC 25922) growth in the presence of ampicillin (10 mg/L) in the microchannels in LB broth at $37^{\circ}$C. (Left) Time-lapse images show that the trapped cells are elongating and swelling, but do not divide, and finally burst in the microchannels. Scale bar, 5 $\mu$m. (Right) The normalized electrical resistance change ${{\Delta R(t)} \over {\Delta R(0)}}$ and the resistance fluctuations of the microchannels as a function of time. Each second in the video corresponds to $\sim3$ min in the experiment.


\bibliography{Supplementary}